\documentstyle[sprocl]{article}
\def\lo{\langle 0 |}
\def\ro{ | 0 \rangle }

 \def\fro{ f_{\rho}}
 \def\mro{m_{\rho}}

\def\gmmu{\gamma _{\mu}}

\def\la{\langle }
\def\ra{ \rangle }

\def\wf{\Psi(\xi,k_{\perp}^{2}) }
\def\wfu{\Psi(u,k_{\perp}^{2}) }
 \def\k{\vec{k}_{\perp}^2}
 
\def\Dm{\vec{iD}_{\mu} }
\newcommand{\beq}{\begin{equation}}
\newcommand{\eeq}{\end{equation}}
\newcommand{\bea}{\begin{eqnarray}}
\newcommand{\eea}{\end{eqnarray}}

\begin{document}

\title{Quark Model,  
 Nonperturbative Wave Functions,  
the QCD Sum Rules and Instantons.}
\author{Ariel~Zhitnitsky\footnote{\it Lectures  
delivered  at the Workshop 
``Nuclear Summer School 97''.
Seoul National  University,  Seoul,
Korea , June 23-June 28 1997.}
}
\address{
        Physics and Astronomy Department \\
        University of British Columbia \\
        6224 Agriculture Road, Vancouver, BC V6T 1Z1, Canada  } 
  \maketitle\abstracts{    
   The main subject of   these lectures  is the Nonperturbative Wave Functions.
We describe  some nonperturbative
methods (like QCD sum rules, dispersion relations, duality etc) in order  
to study this object. We also consider some application of the obtained results, such
as form factors, inclusive amplitudes and diffractive 
electroproduction.
Finally, we discuss the instanton liquid model which
may help us to understand the success of the constituent quark model.   }
                                                
\section{Introduction}
The problem of bound states in the
relativistic quantum field theory  with
large coupling constant is, in general, an extremely 
difficult
problem. Understanding the structure of the
bound state is a very ambitious goal which
assumes the 
solution of a whole spectrum of tightly connected problems,
 such as 
confinement,
chiral symmetry breaking phenomenon,   and many others which
are greatly important in the low energy region. Fortunately, a 
great deal of
 information can be obtained even in absence of such a detailed
 knowledge. 
This happens in high energy processes where the only needed 
nonperturbative 
input are the so-called light-cone wave functions (WF's)
  with a minimal number 
of constituents. 
As is known those WF's     give  parametrically leading contributions 
to hard exclusive processes, hard diffractive electroproduction 
and many others high energy processes.
 In all such cases the quark and antiquark are 
produced at small distances $z\sim 1/Q\rightarrow 0$, where $Q$
is the typical large momentum transfer. Thus, we can neglect
the $z^2$ dependence of the wave function of the
meson with momentum $p$ and can concentrate on the  
variable  $zp \simeq 1$ which is order of one. Therefore, 
the problem is
drastically simplified in the asymptotic limit and we end up
with the light cone wave function which depends only on one variable
 $\phi(zp, z^2=0) $.    

The corresponding 
wave functions have  been introduced to the theory
 in the late seventies
and  early eighties  \cite{Brod} in order to describe the
exclusive processes in QCD. We refer to the review papers 
\cite{CZ},\cite{Brod1}, \cite{Cher1}
 on this subject for the  detail definitions
and discussions in the given context. 
The main idea of the approach \cite{Brod}
 is the separation of the large and small
distance physics. At small distances we can use the standard
perturbative expansion due to the asymptotic freedom and
smallness of the coupling constant. All nontrivial,
large distance physics is hidden into the
nonperturbative   WF  in this approach.
It can not be found
by perturbative technique, but rather  should be extracted
from  elsewhere.   The most powerful analytical
nonperturbative method
for such problems is the
QCD sum rules \cite{SVZ}, \cite{Shif2}.
Therefore,    problem becomes tractable within existing 
nonperturbative methods which are based on the so-called
Wilson Operator Product Expansion (OPE).
    Therefore, the problem is reduced to the 
analysis of the bound states within OPE.
 The   subject of the {\bf Section 2} is   
the corresponding analysis of
the light cone WF's within QCD.

We have therefore a well-formulated problem of the high energy behavior.  
The formulation is based on the solid background of QCD.
However, calculations of the pQCD approach refer, strictly 
speaking, only to
 asymptotically
 high  energies.  Therefore, as usual,
the {\bf main question } remains: at what energies do the 
asymptotic formulae  
  start to work? 

This problem has been a controversial issue for the last fifteen years.  
In order to answer on the question formulated above, we should know
the power corrections to an amplitude under consideration.
It is commonly believed that the power corrections are related to 
the transverse momentum distribution in a hadron.
 Therefore, any dependence on $\k$ gives
some power corrections to the leading terms.
Naively one may  expect that these corrections should be
 small
enough already in the   few $GeV^2$ region.
However, this is not the case. In the {\bf Section 3} we develop 
a theory of the transverse momentum distribution 
 in a hadron.  This is a key element of the whole
approach. The answer on the question about $\k$
will give us the    answer on the formulated above question 
regarding the onset of the asymptotic regime. 

The subject of the {\bf Section 4} is an application of the 
these, apparently pure academic results to
 the phenomenological needs of the particle physics.
Namely, we shall see that the onset of the asymptotic regime
is very different for different processes in spite of the fact that
WF's  which come into the game are always the same.
 
It is instructive to recall a  history of this development
to the problem  about an applicability of the
QCD-based approach to exclusive amplitudes.

As we already mentioned,  the dependence on $\k$ gives
some power corrections to the asymptotically leading terms.
The expectation that these corrections should be
 small
enough already in the   few $GeV^2$ region
  is mainly based 
not on a theoretical analysis but rather on the
 phenomenological observation
that  the dimensional
counting rules,  proposed in early seventies, see 
 \cite{Matveev},
agree well with the experimental data (such as the
 pion and nucleon form factors,
large angle elastic scattering
cross sections and so on). This agreement 
can be interpreted as a strong argument that power corrections 
are small in
 the   few $GeV^2$ region.
 
 However, in mid eighties the applicability of the approach
\cite{Brod}  at experimentally
accessible momentum transfers was questioned \cite{Ditt},
\cite{Isgur}.
In these papers
 it was demonstrated, that the perturbative, asymptotically 
leading contribution is much smaller than the nonleading 
"soft"
one.
Similar conclusion, supporting this result, came from
the different side, from  the QCD sum rules
\cite{Rad},\cite{Smilga}, where the direct
calculation of the form factor has been presented
 at $Q^2\leq 3GeV^2$.
  This method
 has been extended later for 
  larger $  Q^2\leq 10 GeV^2$ \cite{Rad1},\cite{BH}
with the same qualitative result:
the soft contribution is more important in this intermediate
 region
than the leading one.

Therefore, nowadays it is  commonly accepted    that 
the asymptotically leading contribution to the exclusive
 amplitudes
cannot provide
the experimentally observable absolute
values at accessible momentum transfers. 
If we go along with this proposition, then the natural
 question  
 arises: 
 How   can one
explain the very good agreement between the
experimental data and  dimensional counting rules
if the asymptotically leading contribution cannot explain
 the data
for experimentally accessible energies?
A possible answer was suggested recently \cite{ChibZh} 
and can be formulated in the following way: very unusual
 properties
 of the transverse momentum distribution
of a hadron lead to the
{\bf mimicry} of the dimensional counting rules
by the soft  mechanism  
at the extended range of intermediate momentum transfers.
Numerically, the soft term is still more important
than the asymptotically leading contribution at rather
 high 
$Q^2\sim 50\div 100 GeV^2$.  
 
A situation with  hard diffractive electroproduction of the $\rho$ meson
is quite different.
Having the  experience with exclusive processes in mind,
one could expect   a similar behavior
 (i.e. a very slow approach to the asymptotically 
leading prediction)  for
  the  diffractive electroproduction as well.
We shall  argue in the Section 4 that this
 naive 
expectation is {\bf wrong}. 
 The asymptotically leading formula starts to work
 already 
in the region $ Q^2 \simeq 10 \; GeV^2$, where power
 corrections 
do not exceed the $ 20\% $ level in the amplitude. 
 This 
surprising result is a consequence of very special
 properties 
of the WF in the $ \k $ variable (as we mentioned
 earlier, 
the $ \k $ dependence determines corrections to the 
leading term).
    
  In {\bf Section 5} we discuss and formulate some unsolved problems:
  How one can understand the constituent
  quark model in terms of QCD
  (more specifically,  in terms of
   QCD vacuum structure)?
We   discuss the so-called Instanton Liquid Model
\cite{Shu_82a},\cite{Mitya}   which in its present form
correctly reproduces multiple mesonic/baryonic/glueball
 correlation functions,
 and also has an increasing direct support from
 lattice studies of instantons (see for review\cite{Shur}).
We find that many
    properties of the light cone WF's which have been known 
    for a while, can be understood
    from the instanton point of view. 
\section{  $\wf $ and its longitudinal distribution.}
\subsection{General constraints}
The aim of  this section is to provide necessary 
definitions 
and establish some essentially model independent
 constraints on the
 non-perturbative  
 WF  $\wf $ which follow from the use of such
 general methods as dispersion 
relations, duality and large order perturbation theory.

 We define the $\pi$ meson
axial wave function
in the following gauge-invariant way:
\bea
\label{d}
if_{\pi}q_{\mu}\psi_{\pi}(zq,z^2)=
\la 0|\bar{d}(z)\gamma_{\mu}\gamma_5
e^{ig\int_{-z}^z A_{\mu}dz_{\mu}} u(-z)|\pi(q)\ra \\
\nonumber
=\sum_n \frac{i^n}{n!}\la 0|\bar{d}(0)\gamma_{\mu}\gamma_5
(iz_{\nu}\stackrel{\leftrightarrow}{D_{\nu}})^n u(0)|\pi(q)\ra ,
\eea
where
$\stackrel{\leftrightarrow}{D_{\nu}}\equiv
\stackrel{\rightarrow}{D_{\nu}}-\stackrel{\leftarrow}{D_{\nu}}$ and
$\Dm=i\vec{\partial_{\mu}}+gA_{\mu}^a\frac{\lambda^a}{2}$ is the
covariant derivative.
{}From its definition is clear that the set of different
$\pi$ meson matrix elements defines the nonperturbative wave
function $\psi_{\pi}(zq,z^2)$.

The most important part (at  asymptotically high $q^2$) is
the one related to the longitudinal distribution.
In this case $z^2\simeq 0$  the WF
depends only on one $zq$- variable. The corresponding
Fourier transformed
wave function will be denoted as $\phi(\xi)$
and its
  $n-$th moment is given by the
following local matrix element:
\bea
\label{1}
\la 0|\bar{d}\gamma_{\nu}
\gamma_5(i\stackrel{\leftrightarrow}{D_{\mu}}
z_{\mu})^{n} u|\pi(q)\ra=if_{\pi}q_{\nu}
(zq)^{n} \la \xi^{n}\ra=if_{\pi}q_{\nu}(zq)^{n}
\int^1_{-1}d\xi\xi^{n}\phi(\xi)
\eea
\bea
\label{}
-q^2\rightarrow\infty,~~zq\sim 1~~
\xi=2u-1,~~  z^2=0.   \nonumber
\eea
 Therefore, if we knew all matrix elements (\ref{1})
( which are well-defined ) we could restore the whole
 distribution amplitude $\phi(\xi)$.  
In the infinite momentum frame (IMF) $ q_{z}
 \rightarrow \infty $ the distribution amplitude (DA) 
$ \phi(\xi)  $ describes the distribution of the total
 longitudinal momentum $ 
q_{z} $ between the quark and antiquark carrying the
 momenta $ u q_{z} $ and 
$ (1-u) q_{z} $ , respectively. In what follows we
 will use the 
both variables $ 
\xi $ and $ u\; , \;  \bar{u} \equiv  1-u $
 interchangeously. 
 
 The QCD sum rules approach allows one to find the
magnitudes only the few first moments \cite{Cher2}.
As is known, this information is  not enough to reconstruct
the WF; the parametric behavior at $\xi\rightarrow\pm 1$
is the crucial issue in this reconstruction.

To extract the corresponding information,
we use the following duality argument.
Instead of consideration of the  pion DA itself, we study
  the following correlation function with pion quantum numbers:
\bea
\label{2}
i  \int dx e^{iqx}\la 0|T J_{n}^{\|}(x),J_0(0) |0\ra=
(zq)^{n+2}I_{n}(q^2),~~
 J_{n}^{\|}=\bar{d}\gamma_{\nu}z_{\nu}
\gamma_5(i\stackrel{\leftrightarrow}{D_{\mu}}z_{\mu})^{n} u
 \eea
and calculate its asymptotic behavior at large $q^2$.
The result can be presented in the form of the dispersion integral,
whose spectral density is determined by the pure perturbative one-loop
diagram:
\bea
\label{3}
\frac{1}{\pi}\int_0^{\infty} ds\frac{Im I_n^{pert}(s)}{s-q^2},~~
 Im I_{n}(s)^{pert}=\frac{3}{4\pi(n+1)(n+3)}.
 \eea
  We {\bf assume } that the $\pi$ meson gives a
nonzero contribution to the dispersion integral for arbitrary
$n$ and, in particular, for $n\rightarrow\infty$.
Formally, it can be written in the following way
 \bea
\label{3a}
 \frac{1}{\pi}\int_0^{S_{\pi}^n} ds Im I(s)^{pert}_{n}=
\frac{1}{\pi}\int_0^{\infty} ds Im I(s)^{\pi}_{n},
 \eea
   Our assumption means
  that there are no special cancellations
and $\pi$ meson contribution to the
dispersion integral is not zero, i.e. $S_{\pi}^n(\|)\neq 0$,
where we specified the notation for the longitudinal distribution.
In this case at $q^2\rightarrow\infty$ our assumption (\ref{3a})
leads to the following relation:
\bea
\label{4}
  f_{\pi}^2\la \xi^n\ra (n\rightarrow\infty)
\rightarrow\frac{3S_{\pi}^{\infty}(\|)}{4\pi^2n^2}
 \eea
It  unambiguously  implies the  following behavior at
the end-point region \cite{CZ}:
 \bea
\label{5}
 \la\xi^n\ra=\int_{-1}^1d\xi \xi^n\phi(\xi)\sim 1/n^2,~~~~~
   \phi(\xi\rightarrow
\pm 1)\rightarrow (1-\xi^2).
\eea
Few comments are in order.  Because we consider (by definition) the total  
correlation function (\ref{2}), the behavior (\ref{4})
should be fulfilled for any nonperturbative WF no matter what  
the specific shape of WF is.  
 Thus, our first  constraint   looks as follows:
\bea
\label{constraint}
\bullet 1~~~~~~~~~~~~~~~~~~~~~~~~~~~~~~~~~~~~~
   \phi(\xi\rightarrow
\pm 1)\rightarrow (1-\xi^2).
\eea
We want to emphasize that the constraint ( $\bullet 1$)
 is of very general 
origin
and follows directly from QCD. No numerical
 approximations were involved 
in the above derivation. Pre-asymptotic as $ q^2 
 \rightarrow -  \infty $
 perturbative 
and non-perturbative corrections are only able to 
 change the duality  
interval in Eq. (\ref{4}) (which is an irrelevant issue, 
anyhow) but not 
 the parametric $ 1/n^2 $ behavior which remains unaffected.

\subsection{ Numerical constraints in the longitudinal direction }  
We shall discuss here the $\pi$ meson case only. 
Similar consideration can be carried out for other hadrons as well 
within the same technique. We refer 
to review article\cite{CZ} for  the references references.

The QCD sum rules for the moments of the 
$\pi$ meson WF defined above (\ref{1}) have the following form:
\bea
\label{cz}
\frac{1}{\pi}\int_0^{\infty}ds e^{-s/M^2}Im I_n(s)=
\frac{3M^2}{4\pi^2(n+1)(n+3)}+\\  \nonumber
\frac{1}{12M^2}\la\frac{\alpha_s}{\pi}G_{\mu\nu}^2\ra
+\frac{16\pi(11+4n)}{81M^4}\la \sqrt{\alpha_s}\bar{q}q\ra^2+... \\  \nonumber
\frac{1}{\pi}Im I_n(s)= f_{\pi}^2\la\xi^n\ra\delta(s) +\theta(s-s_n)\frac{3}{4\pi^2(n+1)(n+3)},  
\eea
where $M^2$ is so called Borel parameter which varies in the region
where power corrections as well as continuum
contribution (modeled by the standard 
$\theta(s-s_n)$ function) are small enough, i.e. $< 30\%$.
The result of the standard fitting procedure with respect
to the Borel parameter $M^2$ looks as follows:
\bea
\label{2r}
 \la\xi^2\ra_{\mu^2=1.5 GeV^2}=\int_{-1}^1d\xi \xi^2\phi(\xi)\sim 0.38,~~~~
\la\xi^4\ra_{\mu^2=2.2 GeV^2}\sim 0.21
\eea
Now, in order to interpret  these results, we should compare them with 
the corresponding values
of the asymptotic DA $\phi(\xi)^{asym}=3/4(1-\xi^2)$:
\bea
\label{3r}
 \la\xi^2\ra =\int_{-1}^1d\xi \xi^2\phi(\xi)^{asym}= 0.20,~~~~
\la\xi^4\ra =0..086
\eea
For comparison of those DA's we should keep in mind that:\\
a) They both normalized to unity, $\int_{-1}^1d\xi 
 \phi(\xi)^{asym}=\int_{-1}^1d\xi  \phi(\xi)= 1$,\\
b) They both have the same behavior at $\xi\rightarrow \pm1,~~
\phi(\xi\rightarrow
\pm 1)\rightarrow (1-\xi^2)$,\\
c) They have very different few first moments, see eqs (\ref{2r}),
(\ref{3r}).\\
Such a result was a  motivation to suggest the 
so-called `` two-hump" or ``CZ'' 
wave function\cite{Cher2}. Very soon, 
such a WF became a very popular and controversial issue
at the same time, see, e.g. 
\cite{Mikh}$^-$\cite{Rad97}.
  
  It is not the purpose of these lectures 
   to comment  the  corresponding results.
 Rather, I would like to make a remark,
  that there are strong   arguments 
  in favor as well as against of this two-hump WF.
  My opinion is: the true result rests   somewhere in the middle
  of those two numbers: (\ref{2r}) and (\ref{3r}). In different
  words, the QCD sum rule predictions (\ref{2r}) give somewhat
  larger   magnitudes for the moments and asymptotic results (\ref{3r})
  predict somewhat lower values 
  in comparison with what we believe would be  the 
  correct magnitude of the  moments.
  
  Unfortunately the information regarding $\la\xi^2\ra$
  is very difficult to extract by doing a 
  phenomenological analysis of an  experiment. The problem 
  in such of analysis is the following.
  Before   to extract any information from the experiment,
  we have to make   sure that one of the following conditions
is satisfied:\\ 
  {\bf a).} We are already in the asymptotic regime, i.e.
the power corrections are small.
  Therefore, an
  asymptotic formula already works and one can extract a relevant
   information from the corresponding theoretical expression.
 \\
{\bf  b).} We are not in the asymptotic regime. However, the power 
corrections 
 are under control, i.e. we   know the origin of those power corrections
 as well as 
  we know a way of how to estimate them. In this case, again, one could 
extract
 the relevant information regarding the longitudinal distribution $\phi 
(\xi)$.
   
 Unfortunately, none of those conditions is satisfied presently in any 
experiment. 
  Therefore, one can not make a reliable prediction 
 about the  properties of the distribution amplitude $\phi (\xi)$ 
 unless we know precisely  the power corrections.
 That is why an interpretation of an experiment could be very ambiguous
 and very often  it is based on   a strong assumption about power corrections.
 
 We can not improve our understanding by going along the condition 
 a). To satisfy this condition we need to go to the much  higher     energy and momentum transfer
 in comparison to what is available at the moment.
 Therefore, 
 we  shall go along the line b), i.e. we shall try to understand the
  power corrections relevant to the problem. 
 As we mentioned in Introduction, those power corrections 
 are tightly connected to the behavior of 
 a $\wf$ in transverse direction $\k$. Actually it was the main motivation
 for  the recent study of the hadron transverse momentum distribution  
 within QCD\cite{ChibZh},\cite{Zh94},\cite{Zhit2}.
  We present a review of those results in the next section.
  
 \section{Transverse distribution } 
 \subsection{General constraints}
  The  moments in transverse directions  are 
defined analogously to Eq.(\ref{1}) through gauge 
invariant matrix elements
\beq
\label{6}
\la 0|\bar{d}\gmmu\gamma_5
( i\stackrel{\rightarrow}{D_{\nu}}
 t_{\nu})^{2n} u|\pi(q)\ra=if_{\pi}  q_{\mu}
 (-t^2)^n\frac{(2n-1)!!}{(2n)!!}\la 
\vec{k}_{\perp}^{2n} \ra  \;  , 
\eeq
where  transverse vector $t_{\mu}=(0,\vec{t},0)$ 
is perpendicular
 to the hadron momentum $q_{\mu}=(q_0,0_{\perp},q_z)$.
The factor $\frac{(2n-1)!!}{(2n)!!}$ is introduced 
  to
(\ref{6}) to take into account
  the integration over $\phi$ angle in the
 transverse plane:
$\int d\phi (\cos\phi)^{2n}/ \int d\phi= 
{(2n-1)!!}/{(2n)!!}$.
By  analogy with a non-gauge theory we call 
 $\la \vec{k}_{\perp}^{2} \ra$ 
in this equation
  the mean value of the quark perpendicular momentum, 
though  it does not have 
a two-particle interpretation.
Indeed, it
is very different from the naive, gauge dependent  definition like
$\la 0|\bar{d}\gamma_{\nu}
\gamma_5 \partial_{\perp}^2 u|\pi(q)\ra $,
because the physical transverse gluon is  a participant
of this definition.
 We believe that such definition is the useful generalization
of the transverse momentum conception for the interactive quark
system.
Its relation
 to the  higher Fock components will be discussed at
 the end of this section. Its relation to constituent quark model
 and Instanton picture of QCD vacuum will be discussed in Section 5.
 Here we note that Eq. (\ref{6}) is the only possible
 way to define
 $\la \vec{k}_{\perp}^{2} \ra $ in a manner consistent
 with gauge invariance and 
operator product expansion.

To find the behavior  $\la \vec{k}_{\perp}^{2n}
 \ra $ 
at large $ n $ we can 
repeat the previous duality arguments with the
 following
 result
\footnote{ Here
and in what follows we ignore any mild (non-factorial)
 $n$-dependence.}:
\beq
\label{8}
f_{\pi}^2\la \vec{k}_{\perp}^{2n} \ra 
\frac{(2n-1)!!}{(2n)!!}\sim
n!\Rightarrow
 f_{\pi}^2\la \vec{k}_{\perp}^{2n} \ra \sim n!
\eeq
 This behavior   has been
 obtained 
in Ref.\cite{Zhit2}
by  studying of the  large order perturbative series  
for a proper 
correlation function. Dispersion relations 
and duality arguments transform this information
 into Eq.(\ref{8}). 
 It is important to stress that  
any nonperturbative wave function should respect 
 Eq.(\ref{8}) 
 in spite of the fact that
apparently we calculate only  the perturbative part
( see the
comment after Eq.(\ref{5})). The duality turn this
 perturbative information into 
exact properties of the non-perturbative WF.
The most essential feature of Eq. (\ref{8}) is its 
finiteness for arbitrary
 $ n $. This means that higher moments 
\beq
\label{mom}
\la \vec{k}_{\perp}^{2n} \ra =
\int d\k d\xi
\vec{k}_{\perp}^{2n}\Psi(\k, \xi ) 
\eeq
{\bf do exist} for any $ n $. 
In this formula we introduced the non-perturbative
 $ \wf $ normalized to one. 
Its moments are determined by the local
matrix elements (\ref{6})
which are obviously   finite, because they are normalized at 
low normalization point $\mu$.
 The relations to Brodsky and Lepage
notations $\Psi_{BL}(x_1,\vec{k}_{\perp})$
\cite{Brod1} and to the longitudinal distribution
amplitude $\phi(\xi)$ introduced earlier, look as follow:
\beq
\label{9}
\Psi_{BL}(x_1,\vec{k}_{\perp})=\frac{f_{\pi}16\pi^2
}{\sqrt{6}}\wf,~
\int d\k \wf
= \phi(\xi),~\int_{-1}^1 d\xi\phi(\xi)=1
\eeq
where $f_{\pi}=133 MeV$.
The  existence of the arbitrary high moments
$\la \vec{k}_{\perp}^{2n} \ra$ means that the non-perturbative
$\wf$, defined above, falls off at large transverse momentum $\k$
faster than any power function.
   The relation (\ref{8})
fixes the asymptotic behavior of $\wf$ at large $\k$.
Thus, we arrive to the following constraint:
\beq
\label{10}
\bullet 2~~~~~~~~~~~~~~~~
   \la \vec{k}_{\perp}^{2n} \ra =
\int d\k d \xi
\vec{k}_{\perp}^{2n}\wf\sim n!
 ~~~~~n\rightarrow\infty  .
 \eeq
We can now repeat our duality arguments again for an
 arbitrary number of
transverse derivatives and large ($n\rightarrow\infty$)
 number
of longitudinal derivatives. The result reads \cite{Zh94}:

\beq      
\label{11}  
\bullet 3~~~~~~~~~~~~~~~~~~~~~~~~~ 
  \int d\k
\vec{k}_{\perp}^{2k}\Psi(\k, \xi\rightarrow\pm 1 )\sim 
 (1-\xi^2)^{k+1}
\eeq
The constraint (\ref{11}) 
 is extremely important and implies that the $\k$
dependence of   $\Psi(\k, \xi )$ comes
{\bf exclusively in the combination}
$ \k/(1-\xi^2)$ at $\xi\rightarrow\pm 1$.  This means
 that
 the standard assumption
on factorizability  $\Psi(\k, \xi ) =\psi(\k )\phi(\xi)$
 is at variance with very general properties of the theory
 such as duality and 
dispersion relations. The only form of $ \wf $ satisfying 
all the constraints 
(\ref{5}),(\ref{10}) and (\ref{11}) is the Gaussian with a
 very particular argument:
\beq
\label{12}
 \Psi(\k\rightarrow\infty, \xi 
\rightarrow\pm 1)\sim \exp 
\left(-\frac{\k}{\Lambda^2 
(1-\xi^2)} \right)
 \eeq
(here $ \Lambda^2 $ is a mass scale which can be fixed 
by calculating the
 moments $ \la \k \ra , \la \vec{k}_{\perp}^4 \ra $ etc.) 
Strictly speaking, 
so far we have only established the validity of 
Eq.(\ref{12}) in a vicinity 
of the end-point region $ \xi \rightarrow \pm 1 $. 
However, one can argue \cite{Igor} that ,
    the behavior (\ref{12}) can be approximately valid (with some accuracy)
in the whole range  of the $ \xi $ variable.
 
We would like to pause here in order to make the following conjecture.
The Gaussian $\wf$ (reconstructed above from the  QCD analysis)
 not accidentally coincides with the harmonic oscillator
$\wf$ from the constituent quark model.
 We shall discuss this conjecture within the instanton model
 in a more detail in Section5, but   now    let us recall
 some results  from the constituent quark model.

It has been known for a while\cite {Isgur1} that the equal- time (ET)
 wave functions
\beq
\label{qm}
\Psi_{ET}(\vec{q}\,^2)\sim \exp (-\vec{q}\,^2)
\eeq
of the harmonic oscillator in the rest frame
give a very reasonable description of
static meson properties.
Together with Brodsky-Huang-Lepage prescription  
\cite{BHL}
connecting  the equal -time   and the
light-cone wave functions of two  constituents (with mass $m\sim  300 MeV$)
by identification
$$ \vec{q}^2\leftrightarrow\frac{\k+m^2}{4x(1-x)}-m^2,
{}~~\psi_{ET}(\vec{q}^2)\leftrightarrow\psi_{LC}
(\frac{\k+m^2}{4x(1-x)}-m^2),$$
one can reproduce the
Gaussian behavior (\ref{12}) found  from QCD.
It means, first of all, that our identification
of the moments (\ref{6}) defined in  QCD
with the ones defined in quark model,
is the reasonable conjecture.
 The same method can be applied for the analysis
of the asymptotic behavior of the nucleon WF which
in obvious notations takes the form:
\beq
\label{nucl}
\Psi_{nucleon}(\vec{k}_{\perp i}^2
\rightarrow\infty, x_i)\sim \exp(-\sum\frac{ \vec{k}_{\perp i}^2}{x_i}). 
\eeq	
However, there is a difference. In quark model
we do have a parameter which describes the mass
of constituent $m\simeq 300 MeV$. We have nothing like that
in QCD. This difference has very important phenomenological consequences
 which will be discussed   in the next section.
Here we would like to emphasize that there 
is no room for such mass  term in QCD. Its inclusion violates
 the duality constraint
(\ref{5}) since in this case we would have
\beq
\label{qm1}
\la \xi^n \ra  \sim \int_{-1}^{1} d \xi \xi^n \exp 
\left(- \frac{m^2}{\Lambda^2
 (1-\xi^2) } \right)\sim
\exp(-\sqrt{n} ) \; , ~n \rightarrow \infty 
\eeq
instead of the $ 1/n^2 $ behavior (\ref{5}). In other 
words, a true nonperturbative 
WF must respect the asymptotic freedom which is
 incompatible with the quark
 model -type mass term in the WF.  To set this
 more accurately, one 
can say that a possible (scale-dependent) mass term in
 $ \wf $ must renormalize 
to zero at a normalization point $ \mu^2 \sim a \; 
few \; GeV^2 $ where the
 duality arguments apply. This conclusion is not at 
variance with popular
 models for the QCD vacuum such as e.g. the instanton
 vacuum (see \cite{Shur}
 for review). Even more than that: The instanton picture
 does support all properties we have been discussing in 
 these lectures, see
 the section 5 for details.  
   In no sense we claim that a WF like (\ref{12}) exhausts the 
$ \pi$ ( $\rho$) -meson properties, or (\ref{nucl}) exhausts the 
nucleon properties.
  
\subsection{ Lowest moments $ \la \k \ra  \; , \;  
 \la \vec{k}_{\perp}^4 \ra $ 
and vacuum  condensates.}

The general constraints of the previous section are 
insufficient for building up 
a realistic non-perturbative WF for the $\rho$, $\pi$ -mesons or nucleon. To fix 
this, we 
follow the same logic as in the analysis of the distribution
 amplitudes and 
calculate few lowest   $  \la 
\vec{k}_{\perp}^{2n} \ra , n=1,n=2$ moments
 of $ \wf $. The physical meaning of the second moment  
$ \la \k \ra $ is clear:
this quantity serves as a common scale for power corrections 
in physical amplitudes. 
  The fourth moment  $  \la 
\vec{k}_{\perp}^4 \ra $ 
tell us  on how
 strongly $ \wf $ fluctuates in 
the $ \k $ plane. 
In this section we will 
calculate the lowest moments $ \la \k \ra $ and  $ 
 \la \vec{k}_{\perp}^4 \ra $ for $\rho $ and $\pi$ meson $\wf$.
 With the 
knowledge of  $ \la \k \ra $ and $ \la \vec{k}_{\perp}^4 \ra 
$ we then construct a 
model WF $ \wf $ which will be used to estimate higher twist
 effects in different processes in the next section.

 We start our discussion from the 
 the second moment of the $\wf$
in the transverse direction defined by equation
(\ref{6}). For the $\pi$ meson it
was  calculated for the  first time in \cite{CZ} and
independently (using a quite different technique and very different motivation) 
in \cite{Novik}.
Both results are in a full agreement to each other:
\beq
\label{13}
 \la \vec{k}_{\perp}^{2} \ra_{\pi}=\frac{5}{36}\frac{\la
\bar{q}ig\sigma_{\mu\nu}
 G_{\mu\nu}^a\frac{\lambda^a}{2} q \ra }{\la \bar{q}q\ra}
\simeq \frac{5 m_0^2}{36}\simeq (330 MeV)^2 , ~~~m_0^2\simeq 0.8 GeV^2.
\eeq
A very similar calculation can be done for the $\rho$ meson as well\cite{Igor}.
The result looks like this:
\beq
\label{14}
 \la \vec{k}_{\perp}^{2} \ra_{\rho}=(420 MeV)^2   
\eeq
Essentially, the result given by equations 
(\ref{13}) and (\ref{14})  defines the general scale
of all nonperturbative phenomena for the  pion and $\rho$ meson correspondingly.
Those numbers are very much the same.  It is not
accidentally coincides with $ 300\div 400 MeV$  scale which is the
typical magnitude in the hadronic physics.

 To study  the fine properties of the
transverse distribution it is desired to know the next moment.
The problem can be reduced to the analysis of the
 mixed vacuum condensates of dimension seven \cite{Zh94}:
 \beq
\label{15}
  \la\vec{k}_{\perp}^{4}\ra_{\pi} =
 \frac{1}{8}\{   \frac{-3\la \bar{q}g^2\sigma_{\mu\nu}
 G_{\mu\nu} \sigma_{\lambda\sigma}
 G_{\lambda\sigma}q \ra}{4\la\bar{q}q\ra}
+ \frac{ 13\la \bar{q}g^2
 G_{\mu\nu}   G_{\mu\nu}q \ra}{ 9 \la\bar{q}q\ra}\},
\eeq
A similar formula for the $\rho$ meson has been derived in \cite{Igor}
and takes the following form:
\beq
\label{16}
 \la \vec{k}_{\perp}^4 \ra_{\rho} \simeq \frac{3}{10}
 \mro^4 \la u^4 \ra +
 \frac{1}{4}
 \frac{ \lo \bar{q}g^2
 G_{\mu\nu}   G_{\mu\nu}q  \ro }{ \lo
\bar{q} q \ro } - \frac{1}{8} \frac{ \lo  \bar{q}g^2\sigma_{\mu\nu}
 G_{\mu\nu} \sigma_{\lambda\sigma}
 G_{\lambda\sigma}q    \ro }
{ \lo
\bar{q} q \ro }
\eeq
 Therefore,  we have
explicitly expressed
 $  \la \vec{k}_{\perp}^4 \ra $ in terms of the
vacuum expectation values (VEV's)
 of the dimension 7 operators. 
 This is very important result: we essentially say
 that the properties of a hadron are determined by the 
 QCD vacuum structure! One can say even stronger: the
 gluon fluctuations inside of a hadron are essentially nothing, but
 the vacuum fluctuations. They are very strong (they have in general $1 GeV$ scale)
 but   they are approximately the same for all hadrons. 
 This observation gives a chance to understand a constituent quark model:
 all hadrons are build up from   the same constituents which
 originated from QCD vacuum, see section 5 for details.
 
 Coming back to (\ref{15}) and (\ref{16}) 
  one could
 estimate these condensates naively, by 
factorizing them into the products of the quark
 $ \la \bar{\psi} \psi \ra $  and gluon $ \la g^2 G^2 
\ra $ condensates. 
This procedure, based on the factorization hypothesis,
 does not work 
in the given case: there are essential deviations from 
the factorization prediction 
in VEV's (\ref{14},\ref{15}).  The non-factoraziblity of mixed
 quark-gluon matrix 
elements of such type has been studied in \cite{Zhit3,Zh94}
 by two independent
 methods with full agreement in estimates between them. 
The first one
 \cite{Zhit3} was based on the analysis of heavy-light
quark systems,  
while the second method has related the vacuum condensates
 of the form (\ref{14},\ref{15})
 to some pion matrix elements known from PCAC.   Here we only 
formulate the result of this analysis. A measure of  
non-factorizability is
introduced by the correction factors $ K_{1} \;  , 
\; K_{2} $ in the matrix elements
\bea
\label{17}
  \lo \bar{q}g^2
 G_{\mu\nu}   G_{\mu\nu}q  \ro &=& \frac{1}{6} K_{1} 
\lo  g^2
 G_{\mu\nu}   G_{\mu\nu} \ro \;  \lo  
\bar{q} q \ro   \nonumber \\
\lo  \bar{q}g^2\sigma_{\mu\nu}
 G_{\mu\nu} \sigma_{\lambda\sigma}
 G_{\lambda\sigma}q    \ro &=& \frac{-1}{3} K_{2} 
 \lo g^2 G^2 \ro \;  \lo
\bar{q} q \ro 
\eea
( $ K_{1} = K_{2} = 1 $ in the factorization limit). Those
factors are approximately the same $ K_{1} = K_{2} =  K \simeq 3 $ 
for the both mixed 
operators appearing in Eq.(\ref{17}). A possible uncertainty
 of this estimate 
does not exceed 30 \%  \cite{Zh94}. 
In this case, formulae (\ref{16}), (\ref{17}) give the following
estimate:
\bea
\label{17a}
\la \vec{k}_{\perp}^4 \ra_{\rho} \simeq 0.14 GeV^4 .
 \eea
We would like to present the result of these calculation 
for   $ \la \vec{k}_{\perp}^4 \ra_{\rho} $ by
 introducing a 
dimensionless parameter which is quantitative measure of the fluctuations
in the transverse direction:
\beq
\label{18}
R \equiv \frac{  \la \vec{k}_{\perp}^4 \ra_{\rho}}{ \la \k 
\ra_{\rho}^2} \simeq
       4\sim  5  .    
\eeq
For the $\pi$   and $\rho$ mesons parameter $R$ is almost 
the same and it is very large.
It corresponds to an unexpectedly large   hadronic matrix elements
  of   the high dimensional operators.
We have explicitly calculated those m.e. for the operators of dimension seven.
However, we expect that a similar conclusion   also takes place
for arbitrary higher dimensional operators..
 Such a large value of the parameter $R$ corresponds 
 to the strong fluctuations in  the transverse direction.  
 In terms of wave function this property means a very
unhomogeneous distribution in the  transverse direction.
  
  In terms of the
 QCD vacuum structure
 these fluctuations are due to the numerical enhancement of 
the high dimensional
 quark gluon mixed condensates or, what is the same, the 
large magnitude of 
the parameter nonfactorizability $K$ in Eq.(\ref{17}). 
On the microscopical level such  an unhomogeneous distribution
 corresponds to some  fluctuations of the  strong gluon fields
with a small size. 
In this case, factorization prescription (which essentially suggests
a homogeneous distribution of the gluon vacuum fields)
clearly does not work. As a consequence of this, {\bf high dimensional
operators do violate a factorization.}
We identify these  vacuum fluctuations  
with instantons \cite{Shur}, see section 5.

 An additional evidence in favor 
of this view-point is the recent calculation \cite{ShurZh}
of the $\eta'$ matrix element (which can be measured experimentally!)
 of a high-dimensional gluon operator within an
instanton model. 
In general, it is very difficult to have a direct experimental measurement of   
a matrix element of a  high dimensional operator.
In most cases, we use some indirect experiments (see section 4: Applications)
to establish the properties of those matrix elements. Fortunately, 
due to the uniqueness of the $\eta'$ meson a high dimensional operator
can be measured in the CLEO directly!
Such a matrix element    can be explicitly
  extracted \cite{1},\cite{2} from the $B\rightarrow\eta'$ decays\cite{CLEO}.
  This matrix element indeed is very large in agreement with our analysis.
 
 As we shall discuss in   section 5, 
 in the constituent quark model those properties
 correspond to the small size ($1GeV^{-1}$) of the constituent quark 
 in comparison with the hadron size ($1 fm$).

\subsection{Model wave function $ \wf $ }

The results obtained so far are essentially model independent.
 We have fixed 
the form of the high-$\k $ tail of the true nonperturbative WF 
$ \wf $ Eq.(\ref{12}) 
by the use of the quark-hadron duality and dispersion relations. 
Furthermore, we
 calculated the lowest moments $ \la \k \ra $ and $  \la
 \vec{k}_{\perp}^4 \ra $ 
using the equations of motion and QCD sum rules.
Now our purpose is to build some model for the true 
non-perturbative WF $ \wf $ 
which would respect all general constraints of the previous  Sections and 
incorporate the effect
 of strong fluctuations in the transverse $ \k $ plane found 
in the previous 
 section  (see Eq.(\ref{18})).  
 
We start our discussion  from the analysis of the $\wf$
motivated by constituent quark model \cite{Cotanch}, \cite{Isgur1}$^-$
\cite{BHL} (CQM)
\footnote{Here
we neglect all terms
in QCM related to spin part of constituents. In particular,
we do  not consider Melosh transformation and other ingredients
of the light cone$\Leftrightarrow$equal time connection. It does
not effect   any qualitative  results presented here.}.
Such a function is known to give a reasonable description
of static hadron  properties.
The Brodsky-Huang-Lepage prescription \cite{BHL}
 leads to the following form for the pion $\wf$:
 \beq
\label{16a}
 \Psi(\k , u)_{CQM}=A
\exp(-\frac{\k+m^2}{8\beta^2 u\bar{u}} ),
 \eeq
We call this function  as the constituent quark model
 WF. As we already discussed before,  it  satisfies
two general  constraints ($\bullet 2, \bullet 3$), but not  to ($\bullet 1$)
because of the nonzero magnitude for the constituent mass $m$.
We make the standard choice for the parameter $m\simeq 330 MeV$
in accordance with   its physical meaning.   The parameter
$\beta$ can be  determined from the numerical constraint 
for the mean   value $\la\k\ra$ for $\pi$ meson  (\ref{13})
and $\rho$ meson (\ref{14}) correspondingly. They do not differ much and we shall not 
distinguish them for   qualitative discussions. 
 Parameter $A$
is determined by the normalization eq.(\ref{9}).

 To make function    wider in the longitudinal  direction,
    one can insert into formula (\ref{16a})
an additional factor
\beq
\label{19}
(1+g(\mu)[\xi^2-\frac{1}{5}]),~~g(\mu_1)
=g(\mu_2)\cdot(\frac{\alpha_s(\mu_1)}{\alpha_s(\mu_2)})^{50/9b}
\eeq
with additional parameter $g(\mu)$.
This new parameter $g(\mu)$ allows to   adjust $\la\xi^2\ra$
as appropriate. For $\pi$ and $\rho$ meson
WF's this parameter is quite different. For the asymptotic distribution
amplitude   parameter
$g=0$.

Now we go from Constituent Quark Model to    QCD-based WF.
  Before to design  $\Psi_{QCD}(\k,u)$,
let us explain  what do we mean by that.
We define the {\it non-perturbative wave function} $\Psi(\k,u, \mu)_{QCD} $
through its moments which can be expressed in terms of the non-perturbative
matrix elements (\ref{d},\ref{6}). As is known, all non-perturbative matrix elements
are defined in such a way that all gluon's and quark's
virtualities smaller than some parameter $\mu$ (point of normalization)
are hidden in the definition of the `` non-perturbative
matrix elements" (\ref{6}). All virtualities larger than that should be
taken into account explicitly (perturbatively,
due to the asymptotic freedom). In particular, all perturbative tails
like $1/\k$ should be subtracted from  
  the non-perturbative WF by definition.
 The same procedure should be applied
for the  calculation of non-perturbative vacuum condensate $\la
G_{\mu\nu}^2\ra$,
where the perturbative part related to free gluon propagator
 $1/k^2$ is  also  should be subtracted.

With these general remarks in mind
we propose the following form for the
  non-perturbative wave function $\Psi(\k,u, \mu_0)_{QCD} $
at the lowest normalization point:
 \beq
\label{20}
 \Psi(\k , u, \mu_0)_{QCD}=A
\exp(-\frac{\k }{8\beta^2 u\bar{u}} )\cdot\{1+g(\mu_0)[\xi^2-\frac{1}{5}]\}.
 \eeq
 Parameter $\beta$ for this parametrization is found to be
$\beta\simeq$ 0.3 GeV (it corresponds to $R=2.2$ and $\la\k\ra=0.14 GeV^2$).
In comparison with the constituent
quark model the ``only" difference is the absence
 of the mass term $\sim m$ in the exponent. This difference  is a key element.
We
discussed this point 
earlier and we  would like to emphasize 
this point again:   the nonzero mass in $\Psi(\k , u)_{CQM}$
 was  unavoidable part of the wave function within a quark model.
  As we discussed earlier we do not see any room for such  a  term in QCD,  because
its presence   corresponds to the behavior
   \beq
\label{21}
 \la\xi^n\ra=\int_{-1}^1d\xi \xi^n\phi(\xi)
\sim \int_{-1}^1d\xi \xi^n\exp(-\frac{1}{1-\xi^2})\sim
\exp(-\sqrt{n}), ~n\rightarrow\infty, 
  \eeq
which  is in contradiction to  $1/n^2$  result found earlier
(\ref{5}).   
 
One can   implement
the effect of strong $\k$ fluctuations   discussed
 previously and which is quantitatively  expressed in terms
of parameter $R$(\ref{18}). It can be done in a number of ways.
We shall not discuss this point here refering to the original
literature\cite{Igor}. However, we note, that this effect leads to the     
  existence of 
two characteristic scales in $\wf$
    which     has its explanation within  
instanton vacuum liquid model 
 \cite{Shur}. As we already mentioned, such a property of the $\wf$
 could be viewed as an explicit manifestation of the 
 complexity of the hadronic structure: one scale $\sim \; 1 fm$
 determines the hadron size itself; another, new scale
 $\sim 1GeV^{-1}$ determines the hadron substructure, 
 the size of the constituent quark.   
 
 Our conclusion is that the transverse momentum 
distributions for the $\pi$  and 
$ \rho $-mesons are to a large extent alike. 
However, the distribution in the longitudinal direction is very different for those
WF's.

Let us summarize.
We constructed two different types of  wave functions.
The first one, $\Psi_{CQM}$ is motivated by
quark model with its specific mass parameters.
The second type is  motivated by QCD consideration.
 All these WFs have Gaussian behavior at large $\k$.
However, in the case of $\Psi_{CQM}$ this behavior is
related to the nonrelativistic oscillator model,
while for QCD motivated models this behavior
is provided by  constraints discussed in the previous section.

Contrary to the CQM, the QCD motivated wave functions
do not contain the mass parameter $m\simeq 300 MeV$ which is an essential
ingredient of any quark model. Such a term
is absolutely forbidden from the QCD point of view.

 In the next section we   discuss some applications.
In particular, we calculate
the contribution to the pion form factor
caused by these   wave functions.
We shall find  the qualitative difference in behavior
on $Q^2$, which is our main point. We shall also discuss
the behavior of the hard diffractive electroproduction
of $\rho$ meson as a function of  $Q^2$. Again, as in $\pi$ meson 
form factor case we 
shall see a strong influence of the transverse momentum distribution on 
this behavior. We shall also discuss nucleon form factor.
 
\section{Applications}
\subsection{$\pi$ meson form factor}
The first application we consider is the pion form factor. 
 The starting point is the famous Drell-Yan formula
\cite{DY} (for modern, QCD- motivated  employing of this formula,
see \cite{Brod1}),
where the $F_{\pi}(Q^2)$ is expressed in terms of full
wave functions:
\beq
\label{DY}
F_{\pi}(Q^2)=\int\frac{dx d^2\vec{k}_{\perp}}
{16\pi^3}\Psi^*_{BL}
(x ,\vec{k}_{\perp}+(1-x) \vec{q}_{\perp})
\Psi_{BL}(x ,\vec{k}_{\perp}),
\eeq
where $q^2=-\vec{q}_{\perp}^{2}=-Q^2$ is the momentum transfer.
In this formula, the $\Psi_{BL}(x ,\vec{k}_{\perp})$ is
 the full wave function;
the perturbative tail of $\Psi_{BL}(x ,\vec{k}_{\perp})$ behaves as
$\alpha_s/ \k$ for large $\k$ and should be taken into account
explicitly in the calculations. This gives the
 one-gluon-exchange
(asymptotically leading) contribution to the $\pi$ meson form factor in
terms of distribution amplitude $\phi(x)$ \cite{Brod}.

In terms of QCD, the formula (\ref{DY}) is an assumption.
However, 
  we expect, that by taking into account
only  "soft" gluon contribution  (hidden
in the definition of $\k$ (\ref{1})), we catch  the main effect of the soft physics.
There is no proof for that within QCD.
The only argumentation which can be delivered now to support this assumption
is based on the intuitive picture of quark model, where the
current quark and  soft gluons form a constituent quark
 with original quantum numbers.
No evidence for the gluon playing the role of a valence participant
with a finite amount of momentum is seen.

In terms of vacuum structure those soft gluons are nothing but vacuum
fluctuations. Therefore, those gluons are very strong in amplitude 
(they are soft in a sense
of the momentum they carry on which is small).
Those vacuum fluctuations (which are classical
configurations like instantons) 
are much stronger than the quantum fluctuations   carrying a
nonzero momentum.

From the viewpoint of the operator product expansion, the
assumption formulated above, corresponds to the {\bf summing up} a
subset of higher-dimension power corrections. This subset
actually is formed from the infinite number of soft gluons
and  unambiguously
singled out by the definition of non-perturbative $\wf$ (\ref{d}), (\ref{6}).

We refer to the original paper \cite{ChibZh} for details.
Here we formulate the result of calculations.
     The main {\bf qualitative}
difference between  quark model and QCD- motivated wave functions
is as follows: a much slower fall  off at large $Q^2$
is observed for the $\wf$ motivated by QCD.
The qualitative reason for that is the absence of the
mass term, see discussion after the formula (\ref{21}).
Precisely this term
was responsible for the very steep behavior in  
 all previous calculations  based on a  quark model wave function.
 The declining of the form factor getting even slower
if one takes into account the property of the broadening
of  $\wf$ in transverse direction.
  This property corresponds to the
 strong fluctuations in the transverse
direction and quantitatively 
is related to the large  parameter  
 $R$  discussed in the previous section.
  
Therefore, our main  observation is that the QCD based
WF's could
{\bf mimicry}   the dimensional counting rules
by the soft  mechanism  
at the extended range of intermediate momentum transfers.
Numerically, the soft term is still more important
than the asymptotically leading contribution at rather
 high 
$Q^2\sim 50\div 100 GeV^2$.    
     
 \subsection{Nucleon form factor}   
Now we would like to extend our previous analysis
to the nucleon form factor. The starting point,
as before, is the fundamental constraints ($\bullet 1-\bullet 3$),
which being applied to the nucleon wave function imply
the Gaussian behavior with the specific argument (\ref{nucl}).
With these constraints in mind one can model the nucleon
WF in the same way as we did for the pion. Having modeled
the nucleon wave function, one can calculate 
a soft contribution to different
nucleon amplitudes. The corresponding analysis was carried out in the ref.
\cite{Kroll}. Here we quote some results from this paper.

  The most important qualitative result of these calculations
is similar to what we already observed previously in the $\pi$- meson case:
namely, the combination
$Q^4F^{nucl.}(Q^2)$ is almost constant in the extent region of $Q^2$
in spite of the fact that the corresponding  
 ``soft'' contribution   naively  should   
be decreasing function of  $Q^2$.
 The qualitative explanation
of this phenomenon is the same as before and is related to the
absence of the mass term in the QCD- motivated wave function.
   
The next   observation  
  is related to the longitudinal distribution and can be formulated 
as follows: A fit to the different experimental data leads
to a wave function which has  the same type of asymmetry 
 which was  found previously
from the QCD sum rules. The asymmetry is however much more moderate 
numerically than QCD sum rules indicate.
 
  In particular, 
one can calculate the valence quark distribution functions
$u^p(x)$ and $d^p(x)$ at large $x$
 in terms of  the non-perturbative nucleon WF\cite{Kroll}. Two  properties
of the  WF are important to provide such a fit:
The absence of the mass term in the formula (\ref{nucl}) 
(this leads to the correct
power behavior at $x\rightarrow 1$) and a moderate asymmetry
in the longitudinal direction ( this provides an observed ratio for
the $\frac{u^p(x)}{d^p(x)}\simeq 5 ~at~~ x\rightarrow 1$ in the contrast with 
the asymptotic formula prediction which gives value of 2 for the same ratio).

We already mentioned earlier that the   similar conclusion  
is likely to have place for the pion wave function   also.
 Therefore, the general moral, based on already completed calculations,  
  can be formulated in the following way.
 There is a  standard   viewpoint for
the phenomenological success of the dimensional
counting rules: it is based on the
   prejudice  that the leading twist contribution
plays the main role in most cases. This outlook,
as we mentioned earlier, is based on the experimental data, where
the dimensional counting rules work very well.
We suggest here some different explanation for this
phenomenological success.
Our explanation of  the   slow falling off  of the soft contribution
with energy is due to the
specific properties of   non-perturbative $\wf$.
 In particular, we argued that the absence of the
of the mass parameter in the corresponding $\wf$ is the strict QCD constraint.
At the same time this property is responsible for
the  behavior mentioned above. Besides that,
we found a {\bf new} scale ($\sim 1GeV^2$) in the
 problem, in addition to the standard low energy parameter
$\la\k\ra\simeq 0.1 GeV^2$. Both these phenomena 
lead to the  temporarily {\bf mimicry} of the leading twist behavior
in the extent region of $ Q^2 $.  
   
We believe that this  is a new
  explanation of the phenomenological success of the dimensional
counting rules   at available, very modest energies.
Besides that, as we shall see in the last section, the new scale we found
has very natural explanation in terms of the specific vacuum configurations,
instantons\cite{Shu_82a},\cite{Mitya}
\cite{Shur}. They have small size $\sim 1/3 fm$ which is very different from
the hadron size $\sim 1fm$. The last scale is determined by the
density of instantons. The  new $1/3 ~ fm$ scale, we believe, should appear
in the  quark model. However, in the constituent 
quark model this new scale can   be  
nothing, but  the size of a constituent quark, see section 5 for details.
 
\subsection{Hard diffractive electroproduction}

In this section we apply our model WF   for the study
of the pre-asymptotic effect due to the Fermi motion in 
diffractive 
electroproduction of the $\rho$-meson. Our prime goal is
 to get an 
estimate for 
the onset of the asymptotic regime in this problem. 

The applicability of perturbative QCD (pQCD) to the
 asymptotic limit 
of the hard
diffractive electroproduction of vector meson was
 established 
in Ref.\cite{BFGMS} 
using the light-cone perturbation  theory. The 
authors of \cite{BFGMS} have proved that for the
production of 
longitudinally polarized vector mesons by longitudinally
 polarized
 virtual photons the cross section 
can be consistently calculated in pQCD. There was found
 that at 
high $ Q^2 $ the amplitude factorizes in a product of DA's 
of the 
vector meson and virtual 
photon, the light-cone gluon distribution function of a
 target, and a 
perturbatively calculable on-shell scattering 
amplitude of a 
$ q \bar{q} $ pair off the gluon field of the target.
 After 
the factorization of gluons from the 
target the problem seems to be tractable within OPE-like 
methods.

Let us start with the general form of the amplitude as a matrix 
element of the 
electromagnetic current $ j_{\mu} = e ( 2/3 \bar{u} \gmmu u - 
1/3 \bar{d} \gmmu 
d ) $ :
\beq
\label{40}
M = \epsilon_{\mu} \la N(p - r) \rho (q+ r) |  j_{\mu}(q) |
 N(p) \ra
\; , 
\eeq
where $ \epsilon_{\mu} $ is the polarization vector of the
 photon 
(only the longitudinal polarization is considered, see 
\cite{BFGMS})
 and $ r $ stands for the momentum transfer. We will
 neglect the 
masses of the nucleon and 
$ \rho$-meson in comparison with the photon virtuality 
$ Q^2 $ : 
 $ \mro^2 ,~ m_{N}^2 \ll Q^2 $. For the momentum transfer $ r $ 
we consider
 the limit  $ r^2
= 0 $, but $ r_{\mu} \neq 0 $. 
 We next note the following. The factorization 
of the gluon field of the target \cite{BFGMS}, \cite{Strikman}
 means 
that that these 
gluons act as an external field on highly virtual quarks
 produced by
 the photon
with $ Q^2 \rightarrow \infty $.  
 Retaining only the leading contribution, we arrive 
to the 
asymptotic formula of Ref.\cite{BFGMS} in the form   
suggested in 
\cite{Radasym}:
\beq
\label{45}
M = \frac{4 \pi \sqrt{2} e \alpha_{s} \fro}{N_{c}} \frac{1}{Q}
 \int_{0}^{1} 
dX \frac{F_{\xi}^g (X) \sqrt{1-\xi} }{ X(X-\xi + i \delta)} 
\int_{0}^{1}
\frac{\phi(u)}{ u \bar{u} } 
\eeq
where $ \phi(u) $ is the standard $\rho$ meson light-cone DA 
 and $F_{\xi}^g (X)$ is so-called asymmetric gluon distribution
which becomes the usual gluon distribution function $XF_g(X)$
in the symmetric limit $\xi\rightarrow 0$.
 
The most important 
corrections to the asymptotic formula
(\ref{45}) are due to the quark transverse degrees of freedom 
(the Fermi motion). Calculating only this contribution we make an
 educated guess
 on the scale of higher twist corrections in the diffractive 
electroproduction.  
Technically,   this correction can be written in the following form
   \cite{FKS}\footnote{ We follow  the
 notations of
 Ref.\cite{FKS} by reserving the symbol $ T(Q^2) $ for the
 correction 
in the cross section.}:
\beq
\label{46}
\sqrt{T(Q^2)} = Q^4 \frac{\int_{0}^{1} \frac{du}{u \bar{u}} 
\int_{0}^{Q^2} \; d^2 \vec{k}_{\perp}   \Psi(u, \vec{k}_{\perp}^2) 
\frac{1}{(Q^2 + \k/ (u \bar{u}))^2} 
\left( 1 - 2 \frac{ \k/( u \bar{u})}{ (Q^2 + \k/( u \bar{u}) } 
\right)      
}{ \int_{0}^{1} \frac{du}{u \bar{u}} \int_{0}^{Q^2} \; d^2 
\vec{k}_{\perp}  \Psi(u, \vec{k}_{\perp}^2)} 
\eeq
where $ \wfu $ was defined earlier in terms of the local matrix 
elements (\ref{mom}). By definition, for the  asymptotically large $ Q^2 
\rightarrow
 \infty $ we have  $ \sqrt{T(Q^2 
\rightarrow\infty )}  = 1 $. Deviations from $ \sqrt{T(Q^2)} = 1 $ 
determine a region of applicability of the asymptotic formula
 (\ref{45}). 

Now we are in position to discuss numerical estimates for the 
correction factor 
(\ref{46}) in order to answer the main question formulated   
above.
The authors of \cite{FKS} have observed that the choice of
 $ \wfu $ in a factorized form $ \wfu = \phi(u) 
\psi( \vec{k}_{\perp}^2 ) $ leads to a very slowly 
raising function $ \sqrt{T(Q^2)} $ which approaches 0.8 at rather
 high $ Q^2 \geq 40 \; GeV^2 $ depending on the model chosen
 for $ \wfu $.    Using the QCD motivated  WF   we see
a very fast approach to the asymptotic in (\ref{46}). The correction 
factor $ \sqrt{T(Q^2)} $ reaches the value 0.8 already at $ 
Q^2 \simeq 10 \; GeV^2 $.   
Such a behavior implies a rather low onset for the asymptotic regime 
for the hard diffractive electroproduction.
 The technical reason for this behavior is quite clear:
 large values
 of   $ \k /(
u \bar{u}) $ are exponentially suppressed by the 
QCD motivated WF   and 
thus these extra  terms in the integral (\ref{46}) 
give a small contribution   for $ Q^2 > 10 \;
 GeV^2 $, see original paper\cite{Igor} for details.    
 
Thus our final conclusion is that the onset of the asymptotic
 regime for 
diffractive electroproduction of the longitudinally 
polarized $\rho$-meson 
is 
approximately $ Q^2 \simeq 10 \; GeV^2 $ where corrections 
due to the 
quark transverse degree of freedom constitute less than
20 \% in the amplitude.  
This can be traced back to the fact that in the case at
 hand the power 
corrections are given by the matrix elements of local 
operators and in fact are 
fixed completely by the independent calculation of the
 moments. This is the 
consequence of the structure of Eq.(\ref{46}) and the 
fact that the WF is 
a function of the single variable  $ \k /(u\bar{u} ) $. Only 
global,
 but not local characteristics of $ \wfu $ in the $ \k $ 
plane are important.
 This situation can be 
confronted with the case of exclusive processes. In the
 most well
 studied problem of 
the pion form factor the asymptotic regime has been found
 to be pushed further
to $ Q^2 \gg 10 \; GeV^2 $ \cite{BH,ChibZh}, where the 
sub-leading "soft"
 contribution is still larger than the leading asymptotic
 one.  
  There are no reasons to expect the
 onsets of 
the asymptotic regime to be similar in the diffractive 
electroproduction
 and exclusive processes. On contrary, they differ 
parametrically in  
$ 1/\alpha_{s} $. Moreover, the explicit calculations
 suggest that the 
asymptotic regime in the $ \rho$-meson diffractive 
electroproduction starts
 already at $ Q^2 \simeq 10 \; GeV^2 $. We stress that
 this conclusion refers 
only to the diffractive electroproduction of the 
longitudinally polarized 
$\rho$-meson, the situation with the transverse polarization
 or diffractive 
charmonium production can be quite different.

\subsection{Concluding remarks}
To conclude this section we would like to make a comment 
regarding  the shape of the $\pi$ meson distribution amplitude.
This issue has been discussing  for  quite a while.
We already mentioned at the   end of     section 2 that
a direct extraction of the corresponding information from 
the experimental data is an extremely difficult (if possible at all)  
problem.  Only indirect analysis is available at the moment.
The point is that the power corrections 
(which are very difficult to estimate) could be very
important at present energies. To be more specific,
and in order to clarify our point,  let us consider 
  amplitude $\gamma\gamma^{*}\rightarrow \pi$
which has received  some attention recently\cite{CLEO1}$^,$
\cite{Krol1}$^,$\cite{Cao}$^,$\cite{Rad97}.
Some theoretical calculations \cite{Krol1} claim that the "two-hamp"   
wave function disagrees with CLEO results; other calculations\cite{Cao}
conclude that either the asymptotic or the "two-hamp" wave function
is sufficient to describe the data. Much more sophisticated
method\cite{Rad97} which avoids any assumptions
regarding a shape of the wave function, gives  a very good description
of the experiment\cite{CLEO1}. However, in spite of the power
and generality of the method 
advocated in ref.\cite{Rad97}, it
does not allow us to extract an information
regarding  the shape of the wave function.
In order to obtain such an information we have to make 
some additional strong assumptions regarding 
the power corrections   (as well as perturbative corrections) within the 
standard scheme
of  PQCD   calculations originated in refs.\cite{Brod}. 
Therefore, our point is that a direct analysis of the available
experimental data   
can not provide an unambiguous information  about
the properties  of the wave functions.  
 
\section{Instantons and the Constituent Quark Model.  }
Let us remind that the purpose of these lectures is twofold.
First, these lectures
 have strong  applied (or  phenomenological) direction.
The main result obtained in this direction
 could be formulated in the following way:
there are 
many theoretical, phenomenological and  experimental evidences 
 that the ``soft'' contribution (being un-leading parametrically) 
nevertheless
can  temporarily {\bf mimic} the leading twist behavior
in the extent region of $ Q^2:~~3 GeV^2\leq Q^2\leq 40 GeV^2 $.
This is due to the 
specific properties of   $\Psi(\k, x)$ we have been discussing at length
in these lectures. 
 Such a mechanism, if it is correct, would be
 an explanation of the phenomenological success of the dimensional
counting rules   at available, very modest energies
for many different processes. At the same time, the same properties 
of   $\Psi(\k, x)$ applied to the hard electroproduction 
predict the onset of the asymptotic regime at the relatively small
$Q^2\simeq 10 GeV^2$.
 We believe that these are very new results 
 and very new understanding which deserve for the further studying.
  
  The second goal is much more theoretical at the present level
  of understanding, but it may become 
  a   profound  phenomenological tool in the
   hadronic physics in future (at least we hope so). We have in mind 
  the QCD-based description of the WF properties within the 
  instanton model\cite{Shu_82a}$^,$\cite{Mitya}$^,$\cite{Shur}. A reader probably
has  noticed that
  almost  in each section\footnote{See, in particular, 
  a discussion after eq. (\ref{qm1}); the end of section 3.2;
  a discussion after eq. (\ref{21}); the end of section 4.2.}
   while  we discuss  a result obtained 
  in a phenomenological  way, we also try to explain
  the same result in terms of the QCD vacuum structure
  which could provide such a behavior. We also have been  trying
to explain
  those results in terms of the  
  specific (but important)   vacuum 
  configurations which   could be responsible for the
   aforementioned properties. We believe that those
     vacuum  configurations are instantons.
   
   Now we would like to collect all relevant information 
   from the previous sections here, in one place,
   in order to formulate our understanding of the quark model
   in terms of the instanton vacuum configurations.
   This presentation unavoidably can  not  be   self-contained, 
   but rather it is a very fragmentary one.
   
\subsection{Instantons. Historical Remarks.} 
The instanton solution as the
classical solution with nontrivial topology
  was invented in 1975
motivated by the discovery of the 't Hooft-Polyakov monopole.
 Shortly after its discovery, the physical meaning of the instanton
as a tunneling event between degenerate classical vacua was understood.
We refer to a pedagogical review\cite{Coleman} for details.
Here we shortly introduce  some  relevant notations
by emphasizing on the physical meaning of the instanton solution.

The starting point is the transition from Minkowski to Euclidean space.
In this case the best tunneling path in gauge theory which connects
topologically different classical vacua is the solution of classical
equation of motion in Euclidean space.  To find these solutions, it is convenient
to exploit the following identity
\beq
\label{i1}
S=\frac{1}{4g^2}\int d^4xG_{\mu\nu}^aG_{\mu\nu}^a=
\frac{1}{4g^2}\int d^4x[\pm G_{\mu\nu}^a\tilde{G_{\mu\nu}^a}+
\frac{1}{2} (G_{\mu\nu}^a\mp \tilde{G_{\mu\nu}^a})^2],
\eeq
where $\tilde{G_{\mu\nu}^a}=1/2\epsilon_{\mu\nu\lambda\sigma}G_{\lambda\sigma}^a$.
Since the first term is a topological charge $Q$ of the configuration and must be an 
integer, see\cite{Coleman}
\beq
\label{i2}
Q= \frac{1}{32\pi^2}\int d^4x   G_{\mu\nu}^a\tilde{G_{\mu\nu}^a}, 
\eeq
one can argue that the action is minimal if the field is self-dual or antiself-dual
$G_{\mu\nu}^a=\pm \tilde{G_{\mu\nu}^a}$. We call these solutions as an  instanton 
or antiinstanton
correspondingly. From (\ref{i1}) we have $S=\frac{8\pi^2|Q|}{g^2}$ for (anti)self-dual fields,
implying that the tunneling probability is $P_{tunneling}\sim
\exp{(-\frac{8\pi^2|Q|}{g^2})}$, where the coefficient in front of the
exponent is determined by a one-loop calculation. 
The corresponding calculations have been completed by 't Hooft in 1976.
We do not need in our discussions  an explicit formula of the 't Hooft calculations,
however we do need to know few general  results which follow from his formula,
see\cite{Coleman} for details:

1. An each instanton is accompanied by $4N_c$ zero modes
($N_c=3$ is a number of colors), which are 
related to the so-called collective coordinates describing the field of the instanton.
Those degrees of freedom are: center of the instanton $x_0$ (4 modes)
size of the instanton $\rho$(1 mode) orientations of the instanton 
in the color space $\Omega_i, (4N_c-5$ modes).

2. Density of instantons $dn_I$, or what is the same, tunneling probability
is divergent at large $\rho$:
\beq
\label{i3}
P \sim dn_Id^4x_0\sim
\exp{(-\frac{8\pi^2|Q|}{g^2(\rho)})}\frac{d^4x_0d\rho}{\rho^5}\sim \frac{d\rho}{\rho^5}
(\Lambda\rho)^b, ~b =11N_c/3=11.
\eeq
Such a result is a consequence  of the unjustified
perturbative calculations of the $\beta$ function
at large size $\rho$. We come back to this point later.

3. An each instanton is accompanied by the fermionic zero mode as a
consequence of the very general so-called Index theorem.
\subsection{Chiral symmetry breaking and instantons.}
It is believed that the one of the most profound features
of QCD, the chiral symmetry breaking phenomenon, can be explained
by instantons. This idea has been under discussion for a quite a while.
The modern approach to the problem has been suggested in \cite{Mitya}, see also
reviews\cite{Shur}. We refer to the original papers on the subject for details, 
 here we formulate the main ideas and results of this study.

As we mentioned above, in the presence of an instanton, there is an exactly
one zero mode. This zero mode is the chiral mode: it is right-handed for
instantons and left-handed for antiinstantons. As we shall see, 
this fact has the profound phenomenological consequences for many aspects
of the theory: a description of the  chiral symmetry breaking phenomenon, 
the formation of the chiral condensate, 
there appearance of a nonzero constituent quark mass and many others.

Suppose we have infinitely separated $I$ (instanton) and $\bar{I}$ (antiinstanton)
which can be thought as two degenerate states with two zero modes.
Like in the standard quantum mechanical calculations,
when we decrease a distance between $I$ and $\bar{I}$, this degeneracy is lifted
through the diagonalization of the hamiltonian. 
As the result, the obtained two new states
 have non-zero eigenvalues $\lambda_{\pm}$
 which are equal to the overlap integral $T_{I\bar{I}}$
between the original states $\lambda_{\pm}=\pm |T_{I\bar{I}}|$.
  When one adds more $I's$ and $\bar{I}'s$, each of them brings
in a would be zero mode. After the diagonalization they get split symmetrically
 with respect to zero point $\lambda=0$.
 Eventually, for an instanton-antiinstanton ensemble one
gets a continuous band spectrum with 
a spectral density $\nu(\lambda)$ which is finite at $\lambda=0$.
Such a spectrum implies a nonzero magnitude for the chiral condensate
  $\la\bar{q}q\ra\sim \pi\nu(\lambda=0)$ \cite{Banks}. 

 Having explained the physical mechanism of the chiral symmetry breaking
phenomenon we proceed with the
discussion of some quantities which are much closer to the
phenomenological needs of the hadronic physics. 
 Specifically, let us consider
the quark propagator in the instanton vacuum we discussed above.
The corresponding formula has been derived in ref.\cite{Mitya}. The result 
of this calculation has the form of a massive propagator with a 
{\bf momentum-dependent} dynamical mass:
\beq
\label{i4}
S(k)=\frac{\not k+iM(k^2)}{k^2+M^2(k^2)}, 
 M(k)=M_0F^2(k),~~M_0\simeq 350 MeV   
\eeq
\beq 
\nonumber
  F(z)=2z[I_0(z)K_1(z)-I_1(z)K_0(z)-\frac{1}{z}I_1(z)K_1(z)], ~~
F(0)=1 \nonumber
\eeq
In this formula $z=1/2k\rho$ ($\rho$ is the size of instanton, $\bar{\rho}=1/3 fm$);
$M(k\rightarrow\infty)\sim\frac{6}{k^3\rho^3}\rightarrow 0$ as it should
due to the asymptotic freedom.
However, at small $k $, the mass is not zero but finite number
$M(k\rightarrow 0)=M_0\simeq 350 MeV$ which is close to the phenomenological
value for the   constituent  quark mass. 
The value of the chiral condensate is also can be calculated in 
this model in a   
similar  way
with the following result
\bea
\label{i5}
\la \bar{q}q\ra= i\int\frac{d^4k}{(2\pi)^4}Tr S(k)=-4N_c\int\frac{d^4k}{(2\pi)^4}
\frac{M(k)}{M(k)^2+k^2}\simeq -(255MeV)^3,
 \eea
which is again very close to the phenomenological magnitude.
We conclude this short historical introduction into instanton physics
with the following general remark: A systematic numerical study of various
correlation functions, amplitudes, hadronic characteristics
 (like charge radius of a hadron, constant $f_{\pi}$, etc) in the instanton
vacuum demonstrates a good agreement with experiments and phenomenology, see
 reviews\cite{Shur}. Similar conclusion has been recently obtained from direct
lattice measurements\cite{Nigele}. Therefore, it is fair to say that
instantons   explain  the basic properties
of light hadrons and large distance dynamics. 
With this optimistic note about instantons  we proceed with the discussion
of our main subject: the light cone wave functions.
\subsection{Low energy: how to formulate the problem?} 
We start from the following simple
 observation: At low energies it is very difficult to 
 define a WF in terms of the original quark and gluon fields.
  The problem is  that infinite number of different
  matrix elements are equally important in such a
  would be definition. All of them are not zero
 and  all of them have a standard  hadronic scale $1 GeV$.
  Therefore, there is no any special reason to select a quark-aniquark
  WF in comparison, let us say, a quark-aniquark and ten -gluon -field WF.
  To be more specific and in order to explain the main point, 
  let us consider few possible choices. One could start 
from the simplest two-particle WF (we consider pseudoscalar 
WF for simplification) which is normalized to the following
matrix element:
 \beq
\la 0|\bar{d}i\gamma_5u|\pi\ra=\frac{f_{\pi}m_{\pi}^2}{m_u+m_d}\simeq
-\frac{\la \bar{u}u\ra+\la \bar{d}d\ra}{f_{\pi}}.
\eeq  
  One could make one step further and add an arbitrary number of gluons
into this definition:
\beq
\label{G}
\la 0|\bar{d}i\gamma_5(ig\sigma_{\mu\nu}
 G_{\mu\nu}^a\frac{\lambda^a}{2})^n u|\pi\ra=
 -\frac{2\la \bar{q}(ig\sigma_{\mu\nu}
 G_{\mu\nu}^a\frac{\lambda^a}{2})^n q\ra }{f_{\pi}},
\eeq    
where in the last equation we have used PCAC to reduce the matrix element
to the vacuum condensates. All those vacuum condensates are not known 
exactly, but it is believed that  they are not zero and have a
normal hadronic scale which is about $1 GeV$. One could 
define a WF, based on these matrix elements (\ref{G})
in the same way as we did in   eq.(\ref{d}): non of them is
 better or worse than others at low energy.
Moral of this exercise is simple: There is no any reason
to start from the simplest WF
at low energy, because an arbitrary number of gluons 
are not suppressed at all. Therefore, all of them
go on the same footing.

\subsection{High energy: twist classification} 
 Situation at high energies, as we explained in Introduction
is quite different: there is a twist classification
which makes possible to select the leading twist WF uniquely.
 Specifically, 
for  the $\pi$ meson WF it is given by eq.(\ref{d}) with the axial Lorentz structure.
All other WFs give parametrically smaller contribution $\sim(1/Q^2)^n$
to any amplitude.
Therefore, we can  neglect all of them except the leading ones, which
by dimensional reasons, are exactly the WFs with a minimal
number of constituents.  In different words, precisely those
  WFs   represent a
constituent nature of the  hadrons.

What happens when we go to the lower and lower energies?
We do not have the complete answer on this question, but
we hope we convinced a reader (see section 5.1) that the most important
vacuum gluon fluctuations   in hadronic low-energy physics are 
the instantons. Once  we accept this, 
we know the answer on the question formulated above:
The most important quantitative change which instantons make
is the complete reconstruction of the relevant degrees of freedom:
gluons have disappeared; instead a constituent quark
with a momentum-dependent mass  appears $M(k)=M_0F^2(k)$(\ref{i4}).

\subsection{Next steps.}
 Having explained the non-perturbative physics which 
is related to the instantons,
  we now in position to  answer on the question (at least qualitatively)
formulated above:
 what happens to the leading twist
WFs  when energy is getting smaller and smaller?
Our proposal for the answer is: 
 All those WFs acquire a momentum-dependent mass term.   
  In particular, $\Psi(\k , u, \mu_0)_{QCD}$ (\ref{20})
takes the form:
 \beq
\label{i6}
 \Psi(\k , u, \mu_0)_{QCD}=A
\exp(-\frac{\k +M^2(\frac{\k}{u\bar{u}})}{8\beta^2 u\bar{u}} )
\cdot\{1+g(\mu_0)[\xi^2-\frac{1}{5}]\},
 \eeq
with function $M(z)$ given by eq.(\ref{i4}).
A similar replacement should be made for all WFs
we have discussed previously. The main feature
of this replacement is clear: At high energies  when
a typical $k$ is the same order of magnitude as  the external $Q$, 
we have $M(\k\rightarrow\infty)\rightarrow 0$. In this region, 
WF (\ref{i6}) transforms into 
  the leading twist  WF given by     formulae (\ref{20}).
At small energies when 
$M(\k)\rightarrow M_0$, our WF (\ref{i6}) transforms into 
the constituent quark model WF given by eq.(\ref{18}). 
Most important question: What happens to a variety of WFs 
we mentioned above (\ref{G}) in such a transition? Where do they go?

Before to answer on this question, 
we should remark that a large magnitude for all hadronic matrix elements
with gluons (similar to (\ref{G})), is related to the strong vacuum fields
(which we identify specifically
with instantons in this section) and not to
a specific properties of a  hadron. In   the case of the $\pi$ meson this relation follows from
the PCAC. For different hadrons a similar relation is not so obvious as for 
pion. Nevertheless, experience with $\rho$ meson shows\cite{Igor}, that 
all gluonic matrix elements  are large and they are 
related to the vacuum condensates, i.e.
to the vacuum fields. If it is so, 
all would be the new WFs   (similar to eq. (\ref{G})) describe nothing, but 
  some   vacuuum fluctuations. 
The main effect related to those vacuum fluctuations
 is  well known: a quark becomes a constituent quark with 
the momentum dependent mass $M(k)$. In different words, vacuum   gluons
are transformed into  different powers of $k$.

Therefore, 
the answer on the formulated above question regarding
the numerous  of WFs is: they transform into the
 only relevant degrees of freedom, the massive strongly interacting
constituent
quarks with dependent on $k$ mass.
 In terms of these constituent quarks, an apparent variety  
of those WFs simply disappears. This effect can be explained 
using an analogy with the harmonic oscillator problem
in the external electric field. As is known,
in this case, the whole problem is reduced to the change
of variable (one should make a  shift of the coordinate on the amount
which is proportional to the external electric field).
It is exactly what happens in our case: variable $\k$ get shifted
on the amount of  $M(\k)$. Instanton fields are not constant fields, 
therefore, the value of a shift $\sim M$ 
  depends on momentum $\k$. This analogy is even
closer to the real situation, because   a Gaussian dependence
which we derived from the QCD earlier (see section 3.1) is exactly the solution
of the Schr$\ddot{o}$dinger equation for the harmonic oscillator
potential. 

Now we can reverse our arguments
in order to  explain   the reason for the matrix elements
(\ref{G}) to be  large. Those gluon fields $G_{\mu\nu}$ which are 
present in the definition
of the hadronic matrix elements  have nothing to do  
with the specific hadronic  state, but rather, they are related to 
some strong vacuum 
(instanton) fluctuations.  In this picture we also understand   
some relations, like  (\ref{18}) which   have been obtained in a pure phenomenological way. 
Namely, the fluctuations in the transverse direction for the different hadrons
alike \footnote{We checked this explicitly only for the $\pi$ and $\rho$ mesons, 
but we believe that a situation will be the same for all hadrons.} and 
numerically those fluctuations are very large. In the picture we suggest
  this phenomenon has its natural explanation: All matrix elements which
describe transverse momentum distribution are related to
 the  vacuum (instanton) fields;
therefore they are the same for any hadron. Secondly, large magnitude for the 
parameter $R$(\ref{18}) is related to the nonfactorizability of the vacuum condensates
which is also has a natural explanation from the instanton point of view.
New scale we discussed in the previous sections is nothing but the size
of the instanton $\bar{\rho}\sim 1/3 fm$ 
in the formula for $M(k)$(\ref{i4}).  This scale  is much smaller than
the quark mass $M_0\simeq 350 MeV$.

To end this section let me remark that WF (\ref{i6}) with
the momentum-dependent mass may help to resolve many long-standing
problems in the intermediate region of the nuclear and low-energy
particle physics. With this hope we conclude our lectures.

\end{document}